\documentclass[prb,twocolumn,aps,showpacs]{revtex4}
\usepackage{graphicx}
\usepackage{bm}

\begin{document}
\date{\today}

\title{Popov approximation for composite bosons in the BCS-BEC crossover}
\author{P. Pieri and G.C. Strinati}
\affiliation{Dipartimento di Fisica, UdR INFM, 
Universit\`{a} di Camerino, I-62032 Camerino, Italy}

\begin{abstract}
Theoretical treatments of the BCS-BEC crossover need to provide as accurate as 
possible descriptions of the two regimes
where the diluteness condition applies, either in terms of the constituent 
fermions (BCS limit) or of the composite bosons which 
form as bound-fermion pairs (BEC limit).
This has to occur via a single fermionic theory that bridges across these two 
limiting representations.
In this paper, we set up successive improvements of the fermionic theory, that 
result into composite bosons described at the level 
of either the Bogoliubov or the Popov approximations for point-like bosons.
This work bears on the recent experimental advances on the BCS-BEC crossover
with trapped Fermi atoms, which show the need for accurate
theoretical descriptions of BEC side of the crossover.
\end{abstract}


\maketitle

\section{Introduction}

Recent experimental advances \cite{exp-gen-cb,exp-BCS-BEC} with ultracold trapped Fermi atoms have produced systems of composite bosons (dimers), where 
pairs of fermions bind due to their strong mutual attraction induced by a Fano-Feshbach resonance \cite{FF}.
Contrary to the case of point-like bosons, for which the internal fermionic structure is immaterial, for the composite bosons the 
internal fermionic structure is relevant to the extent that their binding energy is comparable with the energy and temperature scales involved in the 
experiments.
Theoretical descriptions of the composite bosons should thus take into account not only their overall bosonic structure (associated with their 
center-of-mass motion), but also their composite nature in terms of the degrees of freedom of the constituent fermions.

Any sensible theoretical approach to the BCS-BEC crossover should rely on a \emph{single} theory, that recovers controlled approximations on 
both sides of the crossover and provides a continuous evolution between them.
In a previous paper \cite{PPS-PRB-04}, the approach originally developed by Popov for a weakly-interacting (dilute) superfluid Fermi gas \cite{Popov} was 
improved by including the effects of the Bogoliubov-Anderson mode \cite{Schrieffer} in the diagonal fermionic self-energy.
With this generalization, it was also shown \cite{PPS-PRB-04} that, in the strong-coupling limit of the fermionic attraction, the same theory is able to 
describe a dilute system of composite bosons within the Bogoliubov approximation \cite{FW}. 

Theoretical work on the condensates of dilute trapped Bose gases \cite{DGPS-99,CC-99} has shown that an accurate description of the experimental
data can be obtained by using the so-called Popov bosonic approximation \cite{Popov}.
This approximation leads to improvements over the Bogoliubov approximation, in that it treats the densities of noncondensed and condensed atoms on the same 
footing, at least in the temperature range (below the critical temperature) where the two densities are comparable to each other.
It has been shown that the use of the Popov bosonic approximation is especially relevant to the thermodynamics of trapped Bose atoms \cite{DGPS-99}. 
It leads, in particular, to a downward shift of the critical temperature \cite{GPS-1996}, which has been recently measured with great accuracy \cite{CT-2004}.
The Popov bosonic approximation was originally conceived for a homogeneous gas \cite{Popov}; its extension to an inhomogeneous gas was described
in detail in Ref.~\onlinecite{Griffin-1996}.

Purpose of the present paper is to devise a fermionic approximation for the BCS-BEC crossover in the broken-symmetry phase, which in the strong-coupling 
limit of the fermionic attraction recovers the Popov approximation for the composite bosons. 
We shall arrive to this Popov approximation by successive improvements of the fermionic theory, starting initially from the BCS mean-field approximation
and then including fluctuations over and above it at different levels of sophistication.
The theory described already in Ref.~\onlinecite{PPS-PRB-04} (see also 
Ref.~\onlinecite{APS-PRB-03}) will be recovered as an intermediate step of this process.
It will turn out that the Popov description of the composite bosons on the strong-coupling side of the crossover corresponds to including pseudo-gap effects 
on top of the BCS approximation in the fermionic two-particle Green's function on the weak-coupling side of the crossover.
Improvements of the theory on the strong-coupling side, as to get a more accurate description of the scattering processes between the composite
bosons, will be further considered along the lines of Ref.~\onlinecite{PS-2000}.
The issue to refine the theory on the weak-coupling side, as to include all possible contributions to an imperfect Fermi gas \cite{GMB}, will instead
be the subject of future study.

In the following, the theory of the BCS-BEC crossover will be considered for a homogeneous system.
Extension to trapped gases may be done, in practice, by considering a local-density approximation, whereby the fermionic chemical 
potential $\mu$ is replaced whenever it occurs by the local quantity $\mu(\mathbf{r}) = \mu -V(\mathbf{r})$ which includes the potential $V(\mathbf{r})$ 
at position $\mathbf{r}$ in the trap.
This procedure has been already implemented at the BCS (mean-field) level in Ref.~\onlinecite{PRA-Rapid-03}, as well as with the inclusion of fluctuations 
in Ref.~\onlinecite{PPPS-PRL-04} (leading to the Bogoliubov description of the composite bosons). 

Our approach is built on a many-body Hamiltonian for fermions only, with an effective mutual attraction represented by a point-contact
interaction.
It thus contrasts an alternative approach based on a boson-fermion model \cite{Timmermans-2001,Holland-2001,Griffin-2002}.
It has, in fact, been shown \cite{SPS-2004} that (at least for the broad resonances currently used in most experiments) a point-contact interaction is by 
far sufficient to get an accurate description of the relevant scattering 
properties.

The present paper deals with several kinds of approximations for the BCS-BEC crossover at a formal level.
In this respect, alternative strategies will be proposed which emphasize the different constraints one has to deal with to obtain a satisfactory 
description of the composite bosons in terms of the constituent fermions.
Detailed numerical calculations at the mean-field level and also with the inclusion of pairing fluctuations, which recover the Bogoliubov description of
the composite bosons, have already been presented in Ref.~\onlinecite{PPS-PRB-04}.
Quantitative comparison among the different kinds of additional approximations discussed in the present paper awaits further numerical calculations, to 
be considered separately.

The plan of the paper is as follows.
In Section II, the steps leading to the Bogoliubov description of the composite bosons (discussed already in Refs.~\onlinecite{PPS-PRB-04} and 
\onlinecite{APS-PRB-03}) 
are reproposed, aiming at emphasizing the hierarchy of approximations for the fermions and the composite bosons.
The role played in this context by the gap equation will be specifically discussed.
In Section III, it will be shown how to modify the fermionic theory in order to describe the composite bosons within the Popov approximation.
In both (Bogoliubov and Popov) cases, it will be also discussed how to enlarge the selection of the fermionic many-body diagrams, as to refine the 
description 
of the scattering processes between the composite bosons.
Section IV gives our conclusions.
For the sake of completeness, the Bogoliubov and Popov descriptions for point-like bosons will be briefly recalled in the Appendix.

\section{Bogoliubov description of composite bosons}

As anticipated in the Introduction, we will obtain the Popov approximation for the composite bosons by setting up a scheme of \emph{successive approximations}
for the constituent fermions.
To this end, it will be useful to retrace the essential steps which have previously led to the description of the composite bosons within the Bogoliubov
approximation (cf. Refs.~\onlinecite{PPS-PRB-04} and \onlinecite{APS-PRB-03}).
By this procedure, the interplay between the approximations adopted for the constituent fermions (which emerge especially on the BCS side of 
the crossover) and the ensuing approximations for the composite bosons (which are mostly relevant to the BEC side of the crossover) will become evident.
The material presented in this Section also serves for setting up the notation that will be needed in the following Section, where the Popov approximation
for the composite bosons will be explicitly introduced.

As also mentioned in the Introduction, the present approach is based on a many-body Hamiltonian for fermions only, with the (effective) mutual attraction 
represented by a point-contact potential $v_{0} \,\, \delta (\mathbf{r})$ where $v_{0}$ is a negative constant.
This choice entails a suitable regularization in terms of a cutoff $k_{0}$ in wave-vector space, by taking the constant $v_{0}$ of the form 
(in three dimensions)
\begin{equation}
v_{0} \, = \, - \, \frac{2 \pi^{2}}{m k_{0}} \, - \, \frac{\pi^{3}}{m a_{F} k_{0}^{2}}      \label{v0}
\end{equation}
where $m$ is the fermion mass and $a_{F}$ the scattering length of the associated (fermionic) two-body problem \cite{PS-2000}.
In this way, the relevant information about the two-body problem is fed into the many-body problem.
By letting $k_{0} \rightarrow \infty$ (and thus $v_{0} \rightarrow 0$) eventually, the classification of the (fermionic) many-body diagrams 
gets considerably simplified in the normal phase \cite{PS-2000} as well as in the broken-symmetry phase \cite{APS-PRB-03}, since only specific 
diagrammatic substructures survive in the limit.
\subsection{Fermions within the BCS approximation}

The BCS mean field represents the simplest approximation for the BCS-BEC crossover in the broken-symmetry phase.
As such, it is often taken as a reference approximation for this crossover, at least at low enough temperatures \cite{Leggett-80}.
With this approximation, the fermionic single-particle Green's functions (in Nambu's notation) are obtained by solving the Dyson's equation in the 
form \cite{FW}:
\begin{eqnarray}
\mathcal{G}_{1 1}^{{\rm BCS}}(p) & = & - \mathcal{G}_{2 2}^{{\rm BCS}}(-p) = \mathcal{G}_{0}(p) +  
 \mathcal{G}_{0}(p) (-\Delta) \mathcal{G}_{2 1}^{{\rm BCS}}(p)       \nonumber \\
\mathcal{G}_{1 2}^{{\rm BCS}}(p) & = & \mathcal{G}_{2 1}^{{\rm BCS}}(p) = \mathcal{G}_{0}(p) (-\Delta) 
\mathcal{G}_{2 2}^{{\rm BCS}}(p) \; .\label{BCS-sp-Green-functions}                                 
\end{eqnarray}
Here, $p=(\mathbf{p},\omega_{s})$ is a four-vector with wave vector $\mathbf{p}$ and fermionic Matsubara frequency $\omega_{s}=(2s+1)\pi/\beta$ 
($s$ being an integer and $\beta$ the inverse temperature), $\mathcal{G}_{0}(p) = (i \omega_{s} - \mathbf{p}^{2}/(2m) + \mu)^{-1}$ is the free-fermion 
propagator, and we set $\hbar=k_{B}=1$ throughout.
The isotropic ($s$-wave) BCS gap function $\Delta$ is obtained by solving the equation:
\begin{equation}
\Delta \, = \, - \, v_{0} \, \int \! \frac{d \mathbf{p}}{(2\pi)^{3}} \, \frac{1}{\beta} \sum_{s} \, \mathcal{G}_{12}^{{\rm BCS}}(p) \, .  \label{Delta-BCS-equation}
\end{equation}
The regularization of the contact potential is provided in this context via the relation
\begin{equation}
\frac{m}{4 \pi a_{F}} \, = \, \frac{1}{v_{0}} \, + \, \int \! \frac{d\mathbf{p}}{(2\pi)^{3}} \, \frac{m}{\mathbf{p}^{2}}       \label{aF-vs-v0}
\end{equation}
which defines the scattering length $a_{F}$ and is equivalent to Eq.~(\ref{v0}).
One is led in this way to introduce the dimensionless coupling parameter $(k_{F} a_{F})^{-1}$ (where $k_{F}$ is the Fermi wave vector related to
the density $n$ via $n=k_{F}^{3}/(3\pi^{2})$).
This parameter ranges formally from $-\infty$ in weak coupling to $+\infty$ in strong coupling, although, in practice, the crossover 
occurs for $(k_{F} |a_{F}|)^{-1} \lesssim 1$.

Besides being relevant \emph{per se} to the BCS-BEC crossover, in the present context the BCS gap equation (\ref{Delta-BCS-equation}) serves to
determine the bosonic chemical potential $\mu_{B}$ which enters the Bogoliubov approximation for the composite bosons in the strong-coupling limit
of the fermionic attraction.
This is because the BCS gap equation (\ref{Delta-BCS-equation}) plays an analogous role to the Gross-Pitaevskii equation for point-like bosons 
\cite{PS-PRL-2003}. 
For a homogeneous system, it thus fixes the relation between $\mu_{B}$ and the condensate density $n_{0}$.

To show this, we perform the sum over the Matsubara frequency in Eq.~(\ref{Delta-BCS-equation}) and eliminate $v_{0}$ using Eq.~(\ref{aF-vs-v0}), 
to obtain:
\begin{equation}
\frac{m}{4 \pi a_{F}} \, + \, \int \! \frac{d\mathbf{p}}{(2\pi)^{3}} \, \left[ 
\frac{\tanh(\beta E(\mathbf{p})/2)}{2 E(\mathbf{p})} \, - \,  \frac{m}{\mathbf{p}^{2}} \right] \, = \, 0               \label{BCS-gap-equation}
\end{equation}
where $E(\mathbf{p})=\sqrt{\xi(\mathbf{p})^{2}+\Delta^{2}}$ with $\xi(\mathbf{p})=\mathbf{p}^{2}/(2m) - \mu$ is the BCS quasiparticle dispersion.
In the strong-coupling ($\Delta\ll|\mu|$) limit \cite{footnote-strong-coupling}, one can approximate the hyperbolic tangent in the expression 
(\ref{BCS-gap-equation}) by unity and expand $E(\mathbf{p})^{-1}$ to first order in $(\Delta / |\mu|)^{2}$. 
The resulting integrals can be done analytically, yielding \cite{PPS-PRB-04}:
\begin{equation}
\frac{\Delta^{2}}{4|\mu|} \, \simeq \, 2 \, \left( \sqrt{2 |\mu| \epsilon_{0}} \, - \, 2 |\mu| \right) \, \simeq \, \mu_{B} \,\, . \label{BCS-gap-equation-sc}
\end{equation}
Here, the formal relation $\mu_{B} = 2 \mu + \epsilon_{0}$ between the bosonic and fermionic chemical potentials \cite{PS-PRB-1996} has been used,  
where $\epsilon_{0} = (m a_{F}^{2})^{-1}$ is the binding energy of the fermionic two-body problem.

The condensate density $n_{0}$ for composite bosons is further identified as $|\alpha|^{2}$, where in the BCS approximation 
\begin{eqnarray}
&&\alpha = \int \! \frac{d \mathbf{p}}{(2\pi)^{3}}  \, \phi(\mathbf{p}) \, \frac{1}{\beta} \sum_{s} 
            \, \mathcal{G}_{12}^{{\rm BCS}}(p) \,\, ,  \label{alpha}\\
&&\phi(\mathbf{p}) = \sqrt{\frac{8 \pi}{a_{F}}} \, \frac{1}{\mathbf{p}^{2} + a_{F}^{-2}} \label{psi}
\end{eqnarray}
being the (normalized) wave function of the fermionic two-body problem.
Using the Green's function (\ref{BCS-sp-Green-functions}), one then obtains in the strong-coupling 
limit \cite{APS-PRB-03}:
\begin{equation}
\alpha \, \simeq \, \Delta \,\, \sqrt{\frac{m^{2} a_{F}}{8 \pi}}  \,\, .                                
      \label{alpha-sc}
\end{equation} 
Correspondingly, the gap equation (\ref{BCS-gap-equation-sc}) reduces to the form
\begin{equation}
\mu_{B} \, \simeq \, \left( \frac{4 \pi a_{B}}{m_{B}} \right) \,\, n_{0}
                                  \label{H-P-Bogoliubov}
\end{equation}
where $m_{B} = 2 m$ is the mass of the composite bosons and $a_{B} = 2 a_{F}$ is the value of the scattering length for the residual boson-boson interaction
within the present approximation.
[We shall discuss below how improved descriptions of the boson-boson scattering result into smaller values of the ratio 
$a_{B}/a_{F}$.]
A relation between $\mu_{B}$ and $n_{0}$ similar to (\ref{H-P-Bogoliubov}) holds for point-like bosons within the Bogoliubov approximation.
Its enforcement on the Bogoliubov quasiparticle dispersion leads to a gapless spectrum, as required on general ground by the Hugenholtz-Pines 
theorem \cite{FW}.

The strong modification of the fermionic chemical potential $\mu$ when passing from the weak-coupling BCS limit (where $\mu \simeq \epsilon_{F}$,
$\epsilon_{F} = k_{F}^{2}/(2m)$ being the Fermi energy) to the strong-coupling BEC limit (where $\mu \simeq - \epsilon_{0}/2$), quite generally, 
results by supplementing the BCS gap equation (\ref{BCS-gap-equation}) with the density equation:
\begin{equation}
n = 2 \int \! \frac{d \mathbf{p}}{(2\pi)^{3}} \frac{1}{\beta} \sum_{s} e^{i\omega_{s}\eta} 
\mathcal{G}_{11}^{{\rm BCS}}(p) 
 \label{BCS-density-equation}
\end{equation}
where $\eta = 0^{+}$ and the factor of $2$ accounts for the two equally populated spin components.
In particular, by expanding the BCS single-particle Green's function (\ref{BCS-sp-Green-functions}) to the leading significant order in $\Delta$,
in the strong-coupling limit one obtains:
\begin{equation}
\frac{n}{2} \, \simeq \, \int \! \frac{d \mathbf{p}}{(2\pi)^{3}} \, \frac{1}{\beta} \, \sum_{s} \, e^{i\omega_{s}\eta} \, 
\left[ \mathcal{G}_{0}(p) \, - \, \Delta^{2} \, \mathcal{G}_{0}(p)^{2} \, \mathcal{G}_{0}(-p) \right]  \,\, .            \label{appr-BCS-density-equation}
\end{equation}
The first term on the right-hand side of Eq.~(\ref{appr-BCS-density-equation}) vanishes in the strong-coupling ($\beta \mu \rightarrow - \infty$) limit.
In the second term we instead approximate
\begin{equation}
\int \!\! \frac{d \mathbf{p}}{(2\pi)^{3}} \, \frac{1}{\beta} \, \sum_{s} \, 
\mathcal{G}_{0}(p)^{2} \, \mathcal{G}_{0}(-p) 
\, \simeq \, - \, \frac{m^{2} \, a_{F}}{8 \, \pi}   \,\, ,                                                                    \label{3-G-0-sc}
\end{equation}
so that Eq.~(\ref{appr-BCS-density-equation}) reduces to
\begin{equation}
\frac{n}{2} \, \simeq \, \Delta^{2} \, \frac{m^{2} \, a_{F}}{8 \, \pi} \, \simeq \, |\alpha|^{2} 
\label{appr-BCS-density-final}
\end{equation}
according to Eq.~(\ref{alpha-sc}).
No depletion of the condensate, therefore, occurs within the BCS mean-field approximation. 
Physically, this corresponds to temperatures much smaller than the critical temperature and to a weak residual boson-boson interaction.
To describe the composite bosons at temperatures comparable with the critical temperature, it is then required to include fluctuation corrections
beyond BCS in the fermionic single-particle Green's functions, as it will be discussed in subsection IIC. 

\subsection{Composite-boson propagators within the Bogoliubov scheme}

\begin{figure}
\begin{center}
\includegraphics*[width=8.1cm]{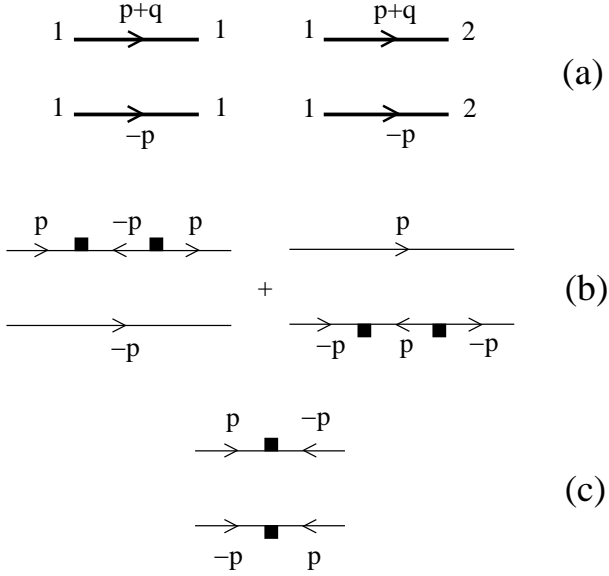}
\caption{(a) Graphical representation of the normal (left) and anomalous 
(right) rungs entering the ladder propagators of Eq.~(\ref{Gamma-propagators}).
(b) Normal and (c) anomalous self-energies for the composite bosons resulting 
from Eqs.(\ref{P-11-approximate}) and (\ref{P-12-approximate}), respectively.
Thick lines stand for the single-particle BCS Green's functions 
(\ref{BCS-sp-Green-functions}), light lines for the free-fermion propagator 
$\mathcal{G}_{0}$, and black squares for (minus) the gap function $\Delta$.} 
\end{center}
\end{figure} 

With the fermions described by the BCS approximation, one can construct the ladder propagators for fermion pairs as described in Ref.\onlinecite{APS-PRB-03}. 
One obtains:
\begin{eqnarray}
\left( \begin{array}{cc} \Gamma_{11}(q) & \Gamma_{12}(q) \\ \Gamma_{21}(q) & \Gamma_{22}(q) \end{array} \right)
\,& =& \, \frac{1}{A(q) \, A(-q) \, - \, B(q)^{2}} \,\,\nonumber\\
&\times&\left( \begin{array}{cc} A(-q) & B(q) \\ B(q) & A(q) \end{array} \right)  \,\, .                                 \label{Gamma-propagators}
\end{eqnarray}
Here, $q=(\mathbf{q},\Omega_{\nu})$ is a four-vector with wave vector $\mathbf{q}$ and bosonic Matsubara frequency
$\Omega_{\nu}=2 \pi \nu/\beta$ ($\nu$ integer), and 
\begin{eqnarray}
- \, A(q) & = & \frac{1}{v_{0}} + \int \! \frac{d \mathbf{p}}{(2\pi)^{3}} \, 
\frac{1}{\beta} \sum_{s}  
                \mathcal{G}_{11}^{{\rm BCS}}(p+q) 
\mathcal{G}_{11}^{{\rm BCS}}(-p) \nonumber \\                                
 B(q) & = &  \int \! \frac{d \mathbf{p}}{(2\pi)^{3}} \, \frac{1}{\beta} 
\sum_{s} \mathcal{G}_{12}^{{\rm BCS}}(p+q) \mathcal{G}_{12}^{{\rm BCS}}(-p)
\label{A-B-definition}
\end{eqnarray}
contain the rungs depicted in Fig.~1(a).
Only these two rungs need, in fact, be considered owing to the regularization (\ref{v0}) of the fermionic potential.
In Ref.~\onlinecite{APS-PRB-03} it was also shown that, in the 
strong-coupling limit of the fermionic attraction, the ladder propagators (\ref{Gamma-propagators}) 
reduce to:
\begin{eqnarray}
& &\Gamma_{11}(q) = \Gamma_{22}(-q) \simeq \frac{8 \pi}{m^{2} a_{F}}  
\frac{\mu_{B} + i \Omega_{\nu} + \mathbf{q}^{2}/(4m)}{E_{B}(\mathbf{q})^{2}  -
 (i \Omega_{\nu})^{2}}\phantom{11111}      \label{Gamma-11-approx}\\
& &\Gamma_{12}(q) = \Gamma_{21}(q)  \simeq  \frac{8 \pi}{m^{2} a_{F}}  
\frac{\mu_{B}}{E_{B}(\mathbf{q})^{2}  -  (i \Omega_{\nu})^{2}}
                  \label{Gamma-12-approx}
\end{eqnarray}
where 
\begin{equation}
E_{B}(\mathbf{q}) =  \sqrt{ \left( \frac{\mathbf{q}^{2}}{2 m_{B}}  +  
\mu_{B}\right)^{2} -  \mu_{B}^{2}} \label{Bogoliubov-disp}
\end{equation}
is the Bogoliubov quasiparticle dispersion with $\mu_{B}$ given by
 Eq.~(\ref{H-P-Bogoliubov}).
The results (\ref{Gamma-11-approx}) and (\ref{Gamma-12-approx}) bear strong resemblance, respectively, with the normal ($G_{B}'$) and anomalous 
($G_{B}^{21}$) noncondensate bosonic Green's functions within the Bogoliubov approximation \cite{FW}.
The comparison gives:
\begin{equation}
\left\{ \begin{array}{l} \Gamma_{11}(q) \, = \, - \, (8 \pi / m^{2} a_{F}) \, G_{B}'(q) \\
\Gamma_{12}(q) \, = \, (8 \pi / m^{2} a_{F}) \, G_{B}^{21}(q) \end{array} \right.                               \label{Gamma-Bogoliubov}
\end{equation}
where the sign difference between the two expressions was accounted for in Ref.~\onlinecite{APS-PRB-03}.

The form (\ref{Gamma-propagators}) of the ladder propagators is obtained by solving the Bethe-Salpeter equation
for the fermionic two-particle Green's function in the particle-particle channel.
It thus holds quite generally for any value of the fermionic coupling $(k_{F} a_{F})^{-1}$.
The rungs in the expressions (\ref{A-B-definition}) contain the BCS single-particle Green's functions which are the \emph{self-consistent} 
solutions of the Dyson's equation (\ref{BCS-sp-Green-functions}). 
This is in accordance with the prescriptions for the approximation to be ``conserving'' \cite{BK-61,Baym-62}. 
In addition, the condition $A(0) - B(0) = 0$, which is required for the ladder propagators (\ref{Gamma-propagators}) to be ``gapless'' at $q=0$, is 
equivalent to the BCS gap equation (\ref{BCS-gap-equation}) for any coupling.
The equivalence can be readily proved via the identity
\begin{equation}
\mathcal{G}_{12}^{{\rm BCS}}(p)  =  \Delta \left[ \mathcal{G}_{11}^{{\rm BCS}}(p)  \mathcal{G}_{11}^{{\rm BCS}}(-p)  + 
\mathcal{G}_{12}^{{\rm BCS}}(p)  \mathcal{G}_{12}^{{\rm BCS}}(-p) \right]
 \label{BCS-identity}
\end{equation} 
which holds for the BCS Green's functions (\ref{BCS-sp-Green-functions}).
It implies, in particular, that the approximation (\ref{Gamma-11-approx}) and (\ref{Gamma-12-approx}) for the propagators of the composite 
bosons is conserving \emph{and} gapless.
This result is consistent with a general property proved in Ref.~\onlinecite{SP-2004}, according to which a given conserving approximation for the constituent
fermions also results into a gapless approximation for the composite bosons (provided the Baym-Kadanoff prescriptions are satisfied also in the two-particle
channel).

In strong coupling the above condition of self-consistency can, however, be relaxed since the system of composite bosons becomes dilute.
The presence of the small ``gas parameter'' $n_{B}^{1/3} a_{B} = r k_{F} a_{F} (6 \pi^{2})^{-1/3}$ (where $n_{B}=n/2$ is the bosonic density and the ratio 
$r = a_{B}/a_{F}$ is of the order unity) suffices, in fact, to select the relevant diagrammatic approximations for a dilute Bose gas (barring the 
temperature regions close to the critical temperature and to zero temperature) \cite{Popov}.

In this case, the forms (\ref{Gamma-11-approx}) and (\ref{Gamma-12-approx}) for the propagators could be obtained by solving a Dyson's equation directly 
for the composite bosons, in an analogous way to Eq.~(\ref{boson-Dyson-equation}) for point-like bosons, with the ladder propagator $\Gamma_{0}(q)$ for 
the normal phase playing the role of the free-boson propagator.
At arbitrary coupling this ladder propagator takes the form \cite{PS-2000}:
\begin{eqnarray}
&&\Gamma_{0}(q) = -  \left\{ \frac{m}{4 \pi a_{F}}  +  
\int \! \frac{d\mathbf{p}}{(2 \pi)^{3}} \right.
\label{most-general-pp-sc}\\ 
&&\!\!\!\times\left[\frac{\tanh(\beta \xi(\mathbf{p})/2) 
+\tanh(\beta \xi(\mathbf{p-q})/2)}{2(\xi(\mathbf{p})+\xi(\mathbf{p-q})
-i\Omega_{\nu})} 
\left. -\frac{m}{\mathbf{p}^{2}} \right] \right\}^{-1}\nonumber      \,\, . 
\end{eqnarray}
In particular, in the strong-coupling limit it reduces to 
\begin{equation}
\Gamma_{0}(q) \simeq  -  \frac{8 \pi}{m^{2} a_{F}} \, \frac{1}{i \Omega_{\nu} 
- \mathbf{q}^{2}/(4m) + \mu_{B}}  \,\, ,    \label{Gamma-o-approx}
\end{equation}
which coincides (apart again from the overall factor $- 8 \pi/(m^{2} a_{F})$) with the free-boson propagator $G_{B}^{(0)}(q)$ of the Appendix.

The corresponding normal and anomalous self-energies for the composite bosons were identified in Appendix B of Ref.~\onlinecite{APS-PRB-03}, by expanding the 
expressions of the rungs (\ref{A-B-definition}), which result after the sum over the Matsubara frequencies is performed, to the leading significant 
order in $\Delta$.
For the following purposes, it is convenient to rephrase that analysis by expanding directly the BCS single-particle Green's functions 
(\ref{BCS-sp-Green-functions}) entering the rungs (\ref{A-B-definition}) to the leading significant order in $\Delta$.
For these rungs one thus obtains:
\begin{eqnarray}
\int \!\! \frac{d \mathbf{p}}{(2\pi)^{3}} \frac{1}{\beta} \sum_{s}  
\mathcal{G}_{11}^{{\rm BCS}}(p+q) \mathcal{G}_{11}^{{\rm BCS}}(-p)
\simeq \int \!\! \frac{d \mathbf{p}}{(2\pi)^{3}}\frac{1}{\beta} \sum_{s} 
 \label{P-11-approximate} \nonumber \\
\times \mathcal{G}_{0}(p+q) \mathcal{G}_{0}(-p) - 2 \Delta^{2} \!\! \int \!\! 
\frac{d \mathbf{p}}{(2\pi)^{3}} 
\frac{1}{\beta} \sum_{s} \mathcal{G}_{0}(p)^{2} \mathcal{G}_{0}(-p)^{2}
\phantom{11}
\end{eqnarray}
and
\begin{eqnarray}
&&\int \!\! \frac{d \mathbf{p}}{(2\pi)^{3}} \frac{1}{\beta} \sum_{s} 
\mathcal{G}_{12}^{{\rm BCS}}(p+q) \mathcal{G}_{12}^{{\rm BCS}}(-p)
\phantom{1111111}\nonumber\\
&& \phantom{1111111}\simeq 
\Delta^{2} \!\! \int \!\! \frac{d \mathbf{p}}{(2\pi)^{3}}
 \frac{1}{\beta} \sum_{s} \mathcal{G}_{0}(p)^{2} 
\mathcal{G}_{0}(-p)^{2}  \label{P-12-approximate}
\end{eqnarray}
with the neglect of the $q-$dependence whenever irrelevant.
Note the presence of a factor $2$ on the right-hand side of Eq.~(\ref{P-11-approximate}).
This factor originates from the two distinct contributions to the normal self-energy for the composite bosons depicted in Fig.~1(b), which result upon dressing 
the upper and lower fermionic lines, respectively.
The anomalous self-energy of Fig.~1(c), on the contrary, contains only one contribution (also with a sign difference with respect to the normal self-energy).

In the expressions (\ref{P-11-approximate}) and (\ref{P-12-approximate}), the quantity
\begin{equation}
\bar{u}_{2}(0,0,0,0) \, \equiv \, \int \! \frac{d \mathbf{p}}{(2\pi)^{3}} \, \frac{1}{\beta} \, \sum_{s} \, 
                              \mathcal{G}_{0}(p)^{2} \, \mathcal{G}_{0}(-p)^{2}                             \label{u-0}
\end{equation}
can be interpreted (apart from an overall factor, see Eq.~(\ref{u-0-sc}) below) as an effective boson-boson interaction where all incoming and outgoing 
four-vectors vanish.
This quantity thus represents a degenerate form of the effective boson-boson interaction for arbitrary four-vectors (depicted in Fig.~2(a))
\begin{figure}
\begin{center}
\includegraphics*[width=8.1cm]{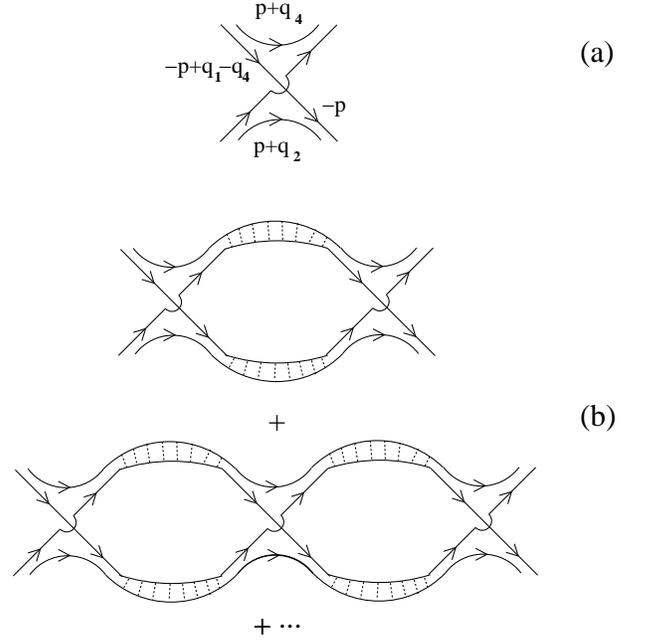}
\caption{(a) Graphical representation of the effective boson-boson interaction
 $\bar{u}_{2}$ of Eq.~(\ref{u-q}).
(b) First additional terms contributed by the t-matrix $\bar{t}_{B}$ for the 
composite bosons given by Eq.~(\ref{bosonic-t-matrix}).
Like in Fig.~1, light lines stand for the free-fermion propagator, while broken
lines stand for the fermionic interaction potential. 
For simplicity, spin labels are not shown explicitly.}
\end{center}
\end{figure} 
\begin{eqnarray}
&&\bar{u}_{2}(q_{1},q_{2},q_{3},q_{4}) \equiv \int \!\! 
\frac{d\mathbf{p}}{(2\pi)^{3}}\frac{1}{\beta} \sum_{s} \nonumber\\
&& \times\,  
\mathcal{G}_{0}(-p) \mathcal{G}_{0}(p+q_{2}) \mathcal{G}_{0}(-p+q_{1}-q_{4}) 
\mathcal{G}_{0}(p+q_{4}) \label{u-q}\phantom{1111}
\end{eqnarray}
with $q_{1}+q_{2}=q_{3}+q_{4}$, which was introduced in Refs.~\onlinecite{PS-PRB-1996} and \onlinecite{PS-2000}.
Note that the expression (\ref{u-q}) contains the free-fermion propagator
 $\mathcal{G}_{0}$ of the normal phase (with an appropriate value of the fermionic 
chemical potential - see below).
This is consistent with the fact that also for point-like bosons (cf. the Appendix) the effective boson-boson interaction (when associated with a t-matrix) is 
considered to be mildly dependent on temperature and is correspondingly evaluated in the normal phase \cite{Popov}.  
In the strong-coupling limit, one obtains for Eq.~(\ref{u-0}):
\begin{eqnarray}
\int \!\! \frac{d \mathbf{p}}{(2\pi)^{3}} \frac{1}{\beta} \sum_{s} 
\mathcal{G}_{0}(p)^{2}\mathcal{G}_{0}(-p)^{2} \simeq
\int \!\! \frac{d \mathbf{p}}{(2\pi)^{3}} \, \frac{1}{4 \xi(\mathbf{p})^{3}}
\nonumber\\
 \simeq \left( \frac{m^{2} a_{F}}{8 \pi} \right)^{2} 
\left( \frac{4 \pi a_{F}}{m} \right) \,\, . 
\label{u-0-sc}
\end{eqnarray}
Apart from the overall factor $(m^{2} a_{F}/(8 \pi))^{2}$ (which is needed to compensate the factors $- 8 \pi/(m^{2} a_{F})$ originating from
the expression (\ref{Gamma-o-approx}) associated with the free-boson propagator), the result (\ref{u-0-sc}) is consistent with the residual 
boson-boson interaction entering Eq.~(\ref{H-P-Bogoliubov}) and yields again the value $2$ for the ratio $r = a_{B} / a_{F}$.

The above considerations suggest us how to improve on the Bogoliubov approximation for the composite bosons, by including, in particular, the diagrammatic
contributions which have been shown in Ref.~\onlinecite{PS-2000} to decrease the ratio $r$ from the value $2$ to about $0.75$ in the zero-density limit
\cite{footnote-smaller-r}.
The price one has to pay for this improvement is to give up the self-consistency of the fermionic single-particle Green's functions, which characterizes
the ladder propagators (\ref{Gamma-propagators}).

To this end, we first obtain the ladder propagators $\mathbf{\Gamma}_{B}(q)$ approximately for any value of the fermionic coupling by adopting a 
Dyson's equation of the type (\ref{boson-Dyson-equation}), where now:

\noindent
(i) The propagators $\mathbf{G}_{B}(q)$ of point-like bosons are replaced by the ladder propagators $\mathbf{\Gamma}_{B}(q)$ of composite bosons; 

\noindent
(ii) In the inverse (\ref{inverse-free-boson}) of the free-boson propagator 
$\mathbf{G}_{B}^{(0)}(q)$, the expression 
$\left(i \Omega_{\nu} - \frac{\mathbf{q}^{2}}{2 m_{B}} + \mu_{B}\right)$ is 
replaced by $\Gamma_{0}(q)^{-1}$ as given by 
Eq.~(\ref{most-general-pp-sc}); 

\noindent
(iii) The bosonic self-energy (\ref{Bogoliubov-self-energy}) within the Bogoliubov approximation is replaced by
\begin{equation}
\mathbf{\Sigma}_{B}(q) \, = \, \Delta^{2} \, \left( \begin{array}{cc} 
- 2 \, \bar{u}_{2}(0,q,0,q) &  \bar{u}_{2}(0,0,-q,q)   \\
\bar{u}_{2}(0,0,-q,q) &  - 2 \, \bar{u}_{2}(0,q,0,q)
\end{array} \right)  \,\, . \label{Bogoliubov-self-energy-cb}
\end{equation}
Note that the $q-$dependence of the diagrams of Fig.~1 has been retained in 
the expression (\ref{Bogoliubov-self-energy-cb}).
In the strong-coupling limit, however, this $q-$dependence can be ignored, and the diagonal and off-diagonal components of the self-energy 
(\ref{Bogoliubov-self-energy-cb}) reduce to the expressions given in Eqs.(\ref{P-11-approximate}) and (\ref{P-12-approximate}), in the order.
The ladder propagators $\mathbf{\Gamma}_{B}(q)$ obtained with the above prescriptions (i)-(iii) reduce correspondingly to the strong-coupling expressions
(\ref{Gamma-11-approx}) and (\ref{Gamma-12-approx}).

Similarly to point-like bosons \cite{Popov}, the (bare) boson-boson interaction entering Eq.~(\ref{Bogoliubov-self-energy-cb}) can be conveniently evaluated 
in the normal phase, even though the corresponding self-energy $\mathbf{\Sigma}_{B}$ refers to the broken-symmetry phase.
This is because in the strong-coupling limit the fermionic chemical potential $\mu$ represents the largest energy scale in the problem 
\cite{footnote-strong-coupling}, so that the precise value of the temperature $T$ is irrelevant insofar as $T\ll|\mu|$.

To guarantee that the ladder propagators $\mathbf{\Gamma}_{B}(q)$ are gapless 
at $q=0$ for any value of the fermionic coupling, the condition
\begin{equation}
\Gamma_{0}(q=0)^{-1} - \Sigma_{B}^{11}(q=0) - \Sigma_{B}^{12}(q=0) =  0
\label{HP-Bogoliubov-cb}
\end{equation}
needs to be satisfied.
In the present context, this condition plays the role of the Hugenholtz-Pines 
theorem for point-like bosons \cite{FW}.
It can be shown that the condition (\ref{HP-Bogoliubov-cb}) reduces to the BCS
 gap equation (\ref{Delta-BCS-equation}) both in strong coupling 
(where it is equivalent to the result (\ref{H-P-Bogoliubov})) and in weak 
coupling.
Accordingly, one expects it to approximate the gap equation 
(\ref{Delta-BCS-equation}) reasonably well for all couplings.

The Bogoliubov approximation for the composite bosons can be improved at this 
point by considering the t-matrix $\bar{t}_{B}(q_{1},q_{2},q_{3},q_{4})$ 
for the composite bosons introduced in Ref.~\onlinecite{PS-2000} and obtained by 
solving the integral equation:
\begin{eqnarray}
&&\bar{t}_{B}(q_{1},q_{2},q_{3},q_{4}) =  \bar{u}_{2}(q_{1},q_{2},q_{3},q_{4})
\nonumber\\
&& - \int \!\! \frac{d \mathbf{q}_{5}}{(2\pi)^{3}} \frac{1}{\beta} 
\sum_{\nu_{5}} \bar{u}_{2}(q_{1},q_{2},q_{5},q_{1}+q_{2}-q_{5}) 
\Gamma_{0}(q_{5})\nonumber \\
&& \times \Gamma_{0}(q_{1}+q_{2}-q_{5}) 
\bar{t}_{B}(q_{1}+q_{2}-q_{5},q_{5},q_{3},q_{4})  \,\, . 
\label{bosonic-t-matrix}
\end{eqnarray}
The first few terms obtained by iteration of Eq.~(\ref{bosonic-t-matrix}) are represented in Fig.~2(b) in terms of the constituent fermions.
These diagrams have been shown in Ref.~\onlinecite{PS-2000} to correct the ratio $a_{B}/a_{F}$ from the value $2$ (obtained with the Born approximation for the 
boson-boson scattering) to about $0.75$.
The analysis made in Ref.~\onlinecite{PS-2000} also suggests that inclusion of these diagrams should become immaterial when approaching the crossover 
region $(k_{F} a_{F})^{-1} \approx 0$, since these diagrams correspond to high-order fermionic pairing-fluctuation processes \cite{Iachello}.
To correct for the value of the boson-boson scattering length entering the ladder propagators (\ref{Gamma-11-approx}) and (\ref{Gamma-12-approx})
in strong coupling, it is thus sufficient to replace $\bar{u}_{2}(q_{1},q_{2},q_{3},q_{4})$ in Eq.~(\ref{Bogoliubov-self-energy-cb}) by the 
t-matrix $\bar{t}_{B}(q_{1},q_{2},q_{3},q_{4})$ of Eq.~(\ref{bosonic-t-matrix}).
Correspondingly, the condition (\ref{HP-Bogoliubov-cb}) must be imposed for the composite bosons to be gapless.
In strong coupling, this condition reduces to the result (\ref{H-P-Bogoliubov}) with the modified value of $a_{B}$.

The integral equation (\ref{bosonic-t-matrix}) was originally introduced in Ref.~\onlinecite{PS-2000} for the normal phase (that is, above the critical temperature 
$T_{c}$).
Extension of that equation below $T_{c}$ would introduce infrared divergences owing to the presence of the Goldstone mode \cite{PCDCS-2004}.
For point-like bosons, Popov \cite{Popov} has avoided these problems by considering the t-matrix associated with the two-body problem (that is, 
in the zero-density limit).
In the present context, we may proceed along similar lines while preserving the composite nature of the bosons.
To this end, we may preliminarly solve the integral equation (\ref{bosonic-t-matrix}) at finite density in the normal phase.
For a given value of the coupling $(k_{F}a_{F})^{-1}$, the temperature is correspondingly kept (slightly) above $T_{c}$, where $T_{c}$  and the chemical
potential $\mu$ are calculated according to the fermionic t-matrix approximation of Ref.~\onlinecite{PPSC-2002}.
This value of $\mu$ is eventually inserted in Eq.~(\ref{bosonic-t-matrix}).

What is still missing for a complete treatment of the Bogoliubov approximation for the composite bosons is to account for the depletion of the condensate,
which is absent in the BCS result (\ref{appr-BCS-density-final}). 
This depletion occurs when approaching the crossover region from strong coupling and/or when the temperature is increased toward $T_{c}$.

\subsection{Fermions within BCS plus pairing fluctuations: Inclusion of the depletion of the condensate}

To find a relation between the condensate ($n_{0}$) and noncondensate ($n'$) densities of the composite bosons, yet preserving the relation 
(\ref{H-P-Bogoliubov}) characteristic of the Bogoliubov theory, it was shown in Ref.~\onlinecite{PPS-PRB-04} that the BCS density equation 
(\ref{BCS-density-equation}) needs to be modified as to include pairing-fluctuation corrections, while preserving the BCS form (\ref{BCS-gap-equation}) of 
the gap equation.
To this end, the fermionic Dyson's equation (\ref{BCS-sp-Green-functions}) is replaced by
\begin{eqnarray}
\mathcal{G}_{11}(p) &=& - \mathcal{G}_{22}(-p) = \mathcal{G}_{0}(p) + 
\mathcal{G}_{0}(p) [\Sigma_{11}(p) \mathcal{G}_{11}(p)    \nonumber \\
&&\phantom{111111111111111111111}+ \Sigma_{12}(p) \mathcal{G}_{21}(p)]
\label{fluctuations-sp-Green-functions}\\ 
\mathcal{G}_{12}(p) &=& \mathcal{G}_{21}(p) = \mathcal{G}_{0}(p) 
[\Sigma_{11}(p) \mathcal{G}_{12}(p) +  \Sigma_{12}(p) \mathcal{G}_{22}(p)]
 \nonumber\,\, ,                              
\end{eqnarray}
where
\begin{eqnarray}
&&\!\!\!\!\!\Sigma_{11}(p)=-\Sigma_{22}(-p) =  -\!\int \!\! 
\frac{d \mathbf{q}}{(2\pi)^{3}} \frac{1}{\beta} \sum_{\nu}  
\Gamma_{11}(q) \mathcal{G}_{11}^{{\rm BCS}}(q-p)\nonumber\\
&&\!\!\!\!\!\Sigma_{12}(p)= \Sigma_{21}(p) = - \Delta 
\label{fermionic-fluctuations-self-energies}                                 
\end{eqnarray}
with $\Gamma_{11}$ given by Eq.~(\ref{Gamma-propagators}) for arbitrary coupling.\begin{figure}
\begin{center}
\includegraphics*[width=8.1cm]{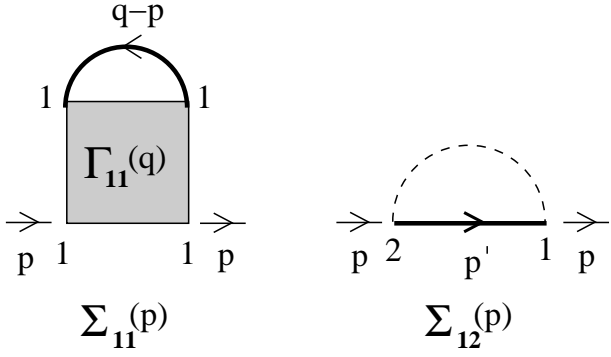}
\caption{Graphical representation of the fermionic self-energies 
(\ref{fermionic-fluctuations-self-energies}) that include pairing fluctuations.
The shaded box corresponds to the normal ladder propagator of 
Eq.~(\ref{Gamma-propagators}). Conventions are like in Figs.1 and 2.}
\end{center}
\end{figure} 
The fermionic self-energies (\ref{fermionic-fluctuations-self-energies}) correspond to the diagrams of Fig.~3. 
The function $\mathcal{G}_{11}^{{\rm BCS}}$ entering Eq.~(\ref{fermionic-fluctuations-self-energies}) has the functional form of the BCS single-particle
Green's function (\ref{BCS-sp-Green-functions}) albeit with modified values of $\Delta$ and $\mu$, different from those obtained (at any given coupling) 
within the BCS approximation of subsection IIA.

As anticipated, the two (gap and density) equations which drive the BCS-BEC crossover are treated here on a different footing \cite{PPS-PRB-04}.
The gap equation is still taken of the form (\ref{Delta-BCS-equation}) with the BCS anomalous Green's function $\mathcal{G}_{12}^{{\rm BCS}}$,
while the density equation now contains the dressed normal Green's function $\mathcal{G}_{11}$ of Eq.~(\ref{fluctuations-sp-Green-functions}) and reads:
\begin{equation}
n \, = \, 2 \, \int \! \frac{d \mathbf{p}}{(2\pi)^{3}} \, \frac{1}{\beta} \, \sum_{s} \, e^{i\omega_{s}\eta} \, \mathcal{G}_{11}(p)
 \label{fluctuations-density-equation}
\end{equation}
in the place of Eq.~(\ref{BCS-density-equation}).

The required relation between $n_{0}$ and $n'$ for the composite bosons can be obtained by expanding $\mathcal{G}_{11}$ in 
Eq.~(\ref{fluctuations-density-equation}) up to second order in the self-energies via the Dyson's equation (\ref{fluctuations-sp-Green-functions}):
\begin{eqnarray}
&&\mathcal{G}_{11}(p) = \mathcal{G}_{0}(p) + \mathcal{G}_{0}(p) 
\Sigma_{11}(p) \mathcal{G}_{0}(p)\nonumber\\
&&\phantom{1111}
+\mathcal{G}_{0}(p) \Sigma_{11}(p) \mathcal{G}_{0}(p)  \Sigma_{11}(p) 
 \mathcal{G}_{0}(p)\nonumber \\
&&\phantom{111}-\mathcal{G}_{0}(p)
\Sigma_{12}(p) \mathcal{G}_{0}(-p) \Sigma_{21}(p) 
\mathcal{G}_{0}(p)  + \cdots             \label{G-11-expansion}      
\end{eqnarray}
which corresponds to an expansion in inverse powers of $|\mu|$.
With the expressions (\ref{fermionic-fluctuations-self-energies}) for the self-energies, the density equation (\ref{fluctuations-density-equation}) then
reduces to:
\begin{widetext}
\begin{eqnarray}
\frac{n}{2} \simeq\int \!\! \frac{d \mathbf{p}}{(2\pi)^{3}} 
\frac{1}{\beta} \sum_{s} e^{i\omega_{s}\eta} \mathcal{G}_{0}(p) 
-  \left( \Delta^{2} + \int \!\! \frac{d \mathbf{q}}{(2\pi)^{3}}  
\frac{1}{\beta} \sum_{\nu}  e^{i\Omega_{\nu}\eta} \Gamma_{11}(q) \right)
\int \!\! \frac{d \mathbf{p}}{(2\pi)^{3}} \frac{1}{\beta} \sum_{s} 
\mathcal{G}_{0}(p)^{2} \mathcal{G}_{0}(-p)\nonumber\\ 
 +   \left( \int \!\! \frac{d \mathbf{q}}{(2\pi)^{3}} \frac{1}{\beta} 
\sum_{\nu} e^{i\Omega_{\nu}\eta}  \Gamma_{11}(q) \right)^{2}
\int \!\! \frac{d \mathbf{p}}{(2\pi)^{3}} \frac{1}{\beta} \sum_{s} 
\mathcal{G}_{0}(p)^{3} \mathcal{G}_{0}(-p)^{2} \,\, .
 \label{fluct-density-equation-espansion}
\end{eqnarray}
\end{widetext}
Note that the $q-$dependence of the single-particle Green's function entering the expression (\ref{fermionic-fluctuations-self-energies}) for $\Sigma_{11}$ 
has been neglected.
Correspondingly, $\Gamma_{11}$ is taken of the polar form (\ref{Gamma-11-approx}) which is valid in the strong-coupling ($\beta \mu \rightarrow - \infty$) 
limit.
The first term on the right-hand side of Eq.~(\ref{fluct-density-equation-espansion}) vanishes in this limit.
In the second term, we write
\begin{eqnarray}
&&\int \!\! \frac{d \mathbf{q}}{(2\pi)^{3}} \frac{1}{\beta}  \sum_{\nu}
 e^{i\Omega_{\nu}\eta} \Gamma_{11}(q)\simeq 
 - \frac{8 \pi}{m^{2} a_{F}} \int \!\! \frac{d \mathbf{q}}{(2\pi)^{3}} 
\nonumber\\
&&\phantom{1111111}\times\frac{1}{\beta}  \sum_{\nu}  e^{i\Omega_{\nu}\eta}  G_{B}'(q) =\frac{8 \pi}{m^{2} a_{F}} \,\, n' \label{definition-n'}
\end{eqnarray}
owing to the first of Eqs.(\ref{Gamma-Bogoliubov}) and to the standard definition of the noncondensate density $n'$ for point-like bosons \cite{FW}.
Using further the result (\ref{3-G-0-sc}) which is also valid in strong coupling, as well as the related result
\begin{equation}
\int \! \frac{d \mathbf{p}}{(2\pi)^{3}} \, \frac{1}{\beta} \, \sum_{s} \, \mathcal{G}_{0}(p)^{3} \, \mathcal{G}_{0}(-p)^{2} \, \simeq \,
- \, \frac{3 m^{4} a_{F}^{5}}{64 \pi}   \,\, ,                                   \label{G3-G2}
\end{equation}
the expression (\ref{fluct-density-equation-espansion}) reduces eventually to:
\begin{equation}
\frac{n}{2} \, \simeq \, n' \, \left(1 \, - \, 3 \pi n' \, a_{F}^{3} \right) \, + \, \Delta^{2} \, \frac{m^{2} a_{F}}{8 \pi}   \label{approximate-n'}
\end{equation}
where $n' a_{F}^{3} \ll 1$ in the dilute limit of interest.

There remains to show that the second term on the right-hand side of Eq.~(\ref{approximate-n'}) represents the condensate density $n_{0}$ even when
pairing fluctuations are included, in agreement with the result (\ref{alpha-sc}) that was proved within the BCS approximation.
To this end, in strong coupling we approximate the diagonal self-energy (\ref{fermionic-fluctuations-self-energies}) in the form:
\begin{eqnarray}
\Sigma_{11}(p) & \simeq & 
\left( - \, \int \! \frac{d \mathbf{q}}{(2\pi)^{3}} \, \frac{1}{\beta} \, \sum_{\nu} \,  e^{i\Omega_{\nu}\eta} \, 
                                                                   \Gamma_{11}(q) \right) \, \mathcal{G}_{11}^{0}(-p)    \nonumber \\
& = & \, \Delta_{0}^{2} \,\,\, \frac{1}{i \omega_{s} + \xi(\mathbf{p})}                                                  \label{Sigma-11-appr-sc}
\end{eqnarray}
where we have used the result (\ref{definition-n'}) for $n'$ and set 

\begin{equation}
\Delta_{0}^{2} \, \equiv \, \frac{8 \, \pi \, n'}{m^{2} \, a_{F}} \,\, .                           \label{definition-Delta-0}
\end{equation}

\noindent
With this expression for the self-energy, the fermionic single-particle Green's functions solutions of the Dyson's equation 
(\ref{fluctuations-sp-Green-functions}) are given by:
\begin{eqnarray}
\mathcal{G}_{11}(p) & = & [i \omega_{s} + \xi(\mathbf{p}) -
 \Sigma_{22}(p)]/\{[ i \omega_{s} - \xi(\mathbf{p}) - \Sigma_{11}(p)] 
\nonumber\\
&\times& [i \omega_{s} + \xi(\mathbf{p}) - \Sigma_{22}(p)]
- \Sigma_{12}(p) \Sigma_{21}(p)\}  \nonumber \\
& \simeq & - \frac{i \omega_{s} + \xi(\mathbf{p})}{\omega_{s}^{2} + 
\gamma(\mathbf{p})^{2} - \Delta_{0}^{2}}   \label{G-11-appr-sc}
\end{eqnarray}
and
\begin{eqnarray}
\mathcal{G}_{12}(p) & = & \Sigma_{12}(p)/
\{[i \omega_{s} - \xi(\mathbf{p}) - \Sigma_{11}(p)]\nonumber\\
&\times& 
[i \omega_{s} + \xi(\mathbf{p}) - \Sigma_{22}(p)]
- \Sigma_{12}(p) \Sigma_{21}(p)\}  \nonumber \\
& \simeq & \frac{\Delta}{\omega_{s}^{2} + \gamma(\mathbf{p})^{2}}
\label{G-12-appr-sc}
\end{eqnarray}
with the notation
\begin{equation}
\gamma(\mathbf{p})^{2} \equiv \xi(\mathbf{p})^{2}  +  \Delta^{2} +  2 
\Delta_{0}^{2} \,\, .           
\label{gamma-definition}
\end{equation}
Using at this point the expression (\ref{G-12-appr-sc}) in the definition
\begin{equation}
\alpha = \int \!\! \frac{d \mathbf{p}}{(2\pi)^{3}}  \, \phi(\mathbf{p}) \, \frac{1}{\beta} \sum_{s} \, \mathcal{G}_{12}(p)         \label{alpha-general}
\end{equation}
which generalizes Eq.~(\ref{alpha}) to the present context, we obtain:
\begin{equation}
\alpha \, \simeq \, \frac{\Delta}{2} \, \int \! \frac{d \mathbf{p}}{(2\pi)^{3}} \, \frac{\phi(\mathbf{p})}{\gamma(\mathbf{p})}
       \, \simeq \, \Delta \,\, \sqrt{\frac{m^{2} a_{F}}{8 \pi}} \,\, .                                            \label{alpha-general-approx}
\end{equation}
This expression coincides with the BCS result (\ref{alpha-sc}) to the 
leading order in $\Delta/|\mu|$.

Entering eventually the result (\ref{alpha-general-approx}) into Eq.~(\ref{approximate-n'}) and identifying again $|\alpha|^{2}$ with the condensate 
density $n_{0}$ for the composite bosons, we get the desired relationship
\begin{equation}
\frac{n}{2} \, \simeq \, n_{0} \, + \, n'    \label{approximate-n'-final}
\end{equation}
which enables us to determine the coupling and temperature dependence of $n_{0}$ in terms of that of $n'$ within the Bogoliubov approximation.
Recall that the Bogoliubov result (\ref{H-P-Bogoliubov}) for the bosonic chemical still holds, since we have kept the BCS gap equation (\ref{Delta-BCS-equation})
unchanged.

As a final remark, we mention that the above results can also be obtained when considering the refinements discussed in subsection IIB to get a value smaller 
than $2$ for the ratio $a_{B}/a_{F}$.

\section{Popov description of composite bosons}

As emphasized in the previous Section, the Bogoliubov approximation rests on the relation (\ref{H-P-Bogoliubov}) between $\mu_{B}$ and $n_{0}$.
For point-like bosons this relation results from the Gross-Pitaevskii equation (in the homogeneous case), while for composite bosons it is obtained 
from the BCS gap equation (\ref{BCS-gap-equation}) as shown in subsection IIA.
For this reason, in subsection IIC the BCS anomalous Green's function $\mathcal{G}_{12}^{{\rm BCS}}$ was still utilized in the gap equation, even
though the dressed normal Green's function $\mathcal{G}_{11}$ entered the density equation.

In the Popov approximation for point-like bosons, on the other hand, the chemical potential is given by the relation $\mu_{B} = g (n_{0} + 2 n')$ 
(cf. the Appendix), a result which can also be derived from a generalized version of the Gross-Pitaevskii equation \cite{Griffin-1996}.
To get a similar relation for the composite bosons, it is thus clear that the BCS gap equation (\ref{BCS-gap-equation}) needs to be modified.

To this end, we enter the expression (\ref{G-12-appr-sc}) for the anomalous fermionic single-particle Green's function (that contains 
pairing-fluctuation corrections beyond mean field) into the gap equation
\begin{equation}
\Delta =  -  v_{0} \int \!\! \frac{d \mathbf{p}}{(2\pi)^{3}} \frac{1}{\beta} 
\sum_{s}  \mathcal{G}_{12}(p)
\label{Delta-general-equation}
\end{equation}
which generalizes Eq.~(\ref{Delta-BCS-equation}).
Considering specifically the strong-coupling limit and taking into account the regularization (\ref{aF-vs-v0}) for the contact potential, 
Eq.~(\ref{Delta-general-equation}) becomes:
\begin{equation}
\frac{m}{4 \pi a_{F}} + \int \!\! \frac{d\mathbf{p}}{(2\pi)^{3}} \left[ 
\frac{\tanh(\beta \gamma(\mathbf{p})/2)}{2 \gamma(\mathbf{p})} -  
\frac{m}{\mathbf{p}^{2}} \right] =  0        \label{general-gap-equation}
\end{equation}
which resembles Eq.~(\ref{BCS-gap-equation}) with $E(\mathbf{p})$ replaced by $\gamma(\mathbf{p})$ of Eq.~(\ref{gamma-definition}).
In strong coupling, the hyperbolic tangent can again be approximated by unity, so that an expression similar to (\ref{BCS-gap-equation-sc}) still results 
but now with $\Delta^{2}$ replaced by $\Delta^{2} + 2 \Delta_{0}^{2}$:
\begin{equation}
\frac{\left( \Delta^{2} + 2 \Delta_{0}^{2} \right)}{4|\mu|} \simeq  
2 \left( \sqrt{2 |\mu| \epsilon_{0}} - 2 |\mu| \right) \simeq \mu_{B} \,\, .
\label{general-gap-equation-sc}
\end{equation}
Since the result (\ref{alpha-general-approx}) holds to the leadi<ng order in $\Delta/|\mu|$ even when $\mathcal{G}_{12}$ is dressed by pairing fluctuations,
Eq.~(\ref{general-gap-equation-sc}) can be cast in the form
\begin{equation}
\mu_{B}  \simeq  \left( \frac{4 \pi a_{B}}{m_{B}} \right)  (n_{0}  +  2  n')
 \label{H-P-Popov}
\end{equation}
provided the relation (\ref{definition-n'}) for $n'$ still holds even within the present Popov approximation (as it will be verified below).
Note that Eq.~(\ref{H-P-Popov}) implies again the Born-approximation result $a_{B} = 2 a_{F}$.

The next step to obtain the Popov approximation for the composite bosons is to identify additional diagrams in the fermionic particle-particle channel,
which reproduce in strong coupling the additional (diagonal) bosonic self-energy (\ref{Popov-self-energy}) of the Appendix needed to obtain the Popov 
approximation for point-like bosons.
Let us similarly call $\Sigma_{B}^{\rm{Pop}}(q)_{11}$ and $\Sigma_{B}^{\rm{Pop}}(q)_{22}$ the corresponding self-energies for the composite bosons.
The diagram associated with the dressing of the upper fermionic line in the ladder propagator is depicted in Fig.~4 (a similar diagram can be drawn for 
the dressing of the lower fermionic line).
It corresponds to the expression:
\begin{figure}
\begin{center}
\includegraphics*[width=5.1cm]{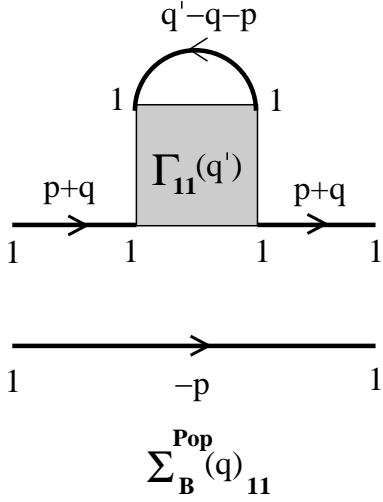}
\caption{Graphical representation of the  bosonic self-energy 
$\Sigma_{B}^{\rm{Pop}}(q)_{11}$ of Eq.~(\ref{Sigma-Popov-comp-bosons-full}), 
obtained by dressing the upper fermionic line in the particle-particle channel.
An analogous dressing can be done for the lower fermionic line. Conventions 
are like in Fig.~3.}
\end{center}
\end{figure} 
\begin{eqnarray}
\Sigma_{B}^{\rm{Pop}}(q)_{11} = - 2 \int \! \frac{d \mathbf{p}}{(2\pi)^{3}} 
 \frac{1}{\beta} \sum_{s}
\int \!\! \frac{d \mathbf{q'}}{(2\pi)^{3}} \frac{1}{\beta} \sum_{\nu'}
\nonumber\\
\times \mathcal{G}_{11}(p+q)^{2} \mathcal{G}_{11}(-p) \mathcal{G}_{11}(q'-q-p) 
\Gamma_{11}(q') \,\, .             
\label{Sigma-Popov-comp-bosons-full}
\end{eqnarray}
[The factor of $2$ in Eq.~(\ref{Sigma-Popov-comp-bosons-full}) accounts for the dressing of the the lower fermionic line, although this is not shown explicitly in Fig.~4.]
In particular, in strong coupling we may approximate the diagram of Fig.~4 as 
follows:
\begin{eqnarray}
&&\Sigma_{B}^{\rm{Pop}}(q)_{11} = \Sigma_{B}^{\rm{Pop}}(-q)_{22} 
  \simeq  - 2  \int \! \frac{d \mathbf{p}}{(2\pi)^{3}} 
\frac{1}{\beta} \sum_{s}\nonumber\\
&&\times\, \mathcal{G}_{0}(p)^{2} \mathcal{G}_{0}(-p)^{2}  
 \int \! \frac{d \mathbf{q'}}{(2\pi)^{3}}  \frac{1}{\beta}  \sum_{\nu'}  e^{i\Omega_{\nu'}\eta}  \Gamma_{11}(q')\nonumber\\           
&& =  -  2  \left( \frac{m^{2} a_{F}}{8 \pi} \right)  \frac{4 \pi a_{F}}{m}  n'
\label{Sigma-Popov-comp-bosons}           
\end{eqnarray}
where $\Gamma_{11}$ is taken of the polar form (\ref{Gamma-11-approx}) as in Eq.~(\ref{fluct-density-equation-espansion}).

The Popov propagators for the composite bosons are now obtained in terms of the corresponding Bogoliubov propagators (\ref{Gamma-propagators}) and of the 
above bosonic self-energy $\Sigma_{B}^{\rm{Pop}}(q)_{11}$, similarly to what is done in Eq.~(\ref{modified-boson-Dyson-equation}) for point-like bosons.
The result is:
\begin{widetext}
\begin{equation}
\left( \begin{array}{cc} \Gamma^{\rm{Pop}}_{11}(q) & \Gamma^{\rm{Pop}}_{12}(q) \\ \Gamma^{\rm{Pop}}_{21}(q) & \Gamma^{\rm{Pop}}_{22}(q) \end{array} \right)
=  \frac{1}{[A(q) - \Sigma_{B}^{\rm{Pop}}(q)_{11}]  [A(-q) - \Sigma_{B}^{\rm{Pop}}(-q)_{11}]  -  B(q)^{2}}                      
\left( \begin{array}{cc} A(-q)  -  \Sigma_{B}^{\rm{Pop}}(-q)_{11} & B(q) \\ B(q) & A(q)  -  \Sigma_{B}^{\rm{Pop}}(q)_{11}\end{array} \right)
 \label{Popov-propagators}
\end{equation}
\end{widetext}
\noindent
where $A(q)$ and $B(q)$ are given by Eq.~(\ref{A-B-definition}).

The propagators (\ref{Popov-propagators}) are gapless provided
\begin{equation}
A(q=0)  - \Sigma_{B}^{\rm{Pop}}(q=0)_{11}  -  B(q=0)  = 0 \,\, .
\label{Gapless-condition-Popov-cb}
\end{equation}
This equation generalizes to the present context the condition $A(q=0)-B(q=0)=0$ for gapless Bogoliubov propagators. 
Note, however, that while the condition $A(q=0)-B(q=0)=0$ is fully equivalent to the BCS gap equation (\ref{BCS-gap-equation}) for any coupling, 
its generalization (\ref{Gapless-condition-Popov-cb}) reduces to the gap equation (\ref{Delta-general-equation}) only in the strong- and weak-coupling limits.
In strong coupling, this can be verified by writing $\gamma(\mathbf{p})^{2}=E(\mathbf{p})^{2}+2\Delta_{0}^{2}$ in Eq.~(\ref{general-gap-equation}) according 
to the definition (\ref{gamma-definition}), by expanding the resulting expression to first order in the small quantity $(\Delta_{0}/E(\mathbf{p}))^{2}$,
and by recalling the expression (\ref{Sigma-Popov-comp-bosons}).
In weak coupling, on the other hand, the fermionic self-energy $\Sigma_{11}(p)$ of Eq.~(\ref{fermionic-fluctuations-self-energies}) can be split into
a constant part $\Sigma_{0}$ and a remainder $\Sigma_{11}^{R}(p)$, which depends explicitly on $p$ and whose magnitude is much smaller than $\Delta$.
By expanding the expression (\ref{fluctuations-sp-Green-functions}) for $\mathcal{G}_{12}$ to the lowest significant order beyond BCS, one obtains in this way:
\begin{eqnarray}
\mathcal{G}_{12}(p)  \simeq  \mathcal{G}_{12}^{{\rm BCS}}(p) 
 -  \Delta  \Sigma_{11}^{R}(p)   \frac{i \omega_{s} + \xi(\mathbf{p})}{(i \omega_{s})^{2} - E(\mathbf{p})^{2}}\nonumber\\ 
 +  \,\Delta  \Sigma_{11}^{R}(-p)  \frac{i \omega_{s} - \xi(\mathbf{p})}{(i \omega_{s})^{2} - E(\mathbf{p})^{2}} 
 \label{Sigma-11-approx-wc}
\end{eqnarray}

\noindent
where the chemical potential has been shifted as $\mu \rightarrow \mu - 
\Sigma_{0}$ whenever it appears.
Correspondingly, the gap equation (\ref{Delta-general-equation}) becomes:      
\begin{eqnarray}
\!\!\!\!\!\frac{1}{v_{0}} &+& \int \! \frac{d \mathbf{p}}{(2\pi)^{3}} 
\frac{\tanh(\beta E(\mathbf{p})/2)}{2 E(\mathbf{p})}\nonumber\\ 
&+& 
2  \int \! \frac{d \mathbf{p}}{(2\pi)^{3}}  \frac{1}{\beta}  \sum_{s}  \Sigma_{11}(p)  \mathcal{G}_{0}(p)^{2}  \mathcal{G}_{0}(-p) 
 =  0  \,\, .                                                                                                             \label{general-gap-equation-wc}
\end{eqnarray}

\noindent
Note that the full $\Sigma_{11}(p)$ has be restored in Eq.~(\ref{general-gap-equation-wc}), since the last term on its right-hand side would vanish
if $\Sigma_{11}(p)$ were taken to be a constant, according to the approximations valid in the BCS limit.
With the aid of Eq.~(\ref{fermionic-fluctuations-self-energies}), in the last term on the right-hand side of Eq.~(\ref{general-gap-equation-wc}) one recognizes 
the weak-coupling limit of the expression (\ref{Sigma-Popov-comp-bosons-full}) for the Popov bosonic self-energy.
The weak-coupling limit of the gapless condition (\ref{Gapless-condition-Popov-cb}) is thus recovered from the gap equation (\ref{Delta-general-equation}).
Recall in this context that the need for including the constant shift $\Sigma_{0}$ in the fermionic chemical potential entering the weak-coupling 
expressions of the rungs (\ref{A-B-definition}) was emphasized in Ref.~\onlinecite{PPS-PRB-04} for all temperatures in the broken-symmetry phase.

At arbitrary coupling, the broken-symmetry requirement for the particle-particle propagator to be gapless can thus be satisfied by imposing the condition 
(\ref{Gapless-condition-Popov-cb}) in the place of the gap equation (\ref{general-gap-equation}).
As shown above, this replacement does not affect quantitatively the weak- and strong-coupling limits of the theory.

Caution must be exerted when implementing the form (\ref{Sigma-Popov-comp-bosons-full}) of the bosonic self-energy, where the bosonic
propagator $\Gamma_{11}$ appears.
This propagator was originally meant to be taken of the form (\ref{Gamma-propagators}), which is valid within the Bogoliubov approximation and to which 
the gapless condition $A(q=0)-B(q=0)=0$ applies.
However, to use the alternative gapless condition (\ref{Gapless-condition-Popov-cb}) for the Popov propagators (\ref{Popov-propagators}) as well as for 
the bosonic self-energy (\ref{Sigma-Popov-comp-bosons-full}) on which these propagators are built, $\Gamma_{11}$ in Eq.~(\ref{Sigma-Popov-comp-bosons-full})
must be interpreted as given by Eq.~(\ref{Popov-propagators}) itself.
This procedure introduces a self-consistency in the bosonic calculation.
By doing so, the above results are valid to the extent that Eq.~(\ref{definition-n'}) holds also within the Popov approximation.

An alternative approach, which does not require reaching self-consistency in the bosonic expressions, would be to maintain the Bogoliubov form 
(\ref{Gamma-propagators}) for the propagator $\Gamma_{11}$ entering the bosonic self-energy (\ref{Sigma-Popov-comp-bosons-full}) with the associated 
gapless condition $A(q=0)-B(q=0)=0$, while imposing the modified gapless condition (\ref{Gapless-condition-Popov-cb}) on the Popov propagators 
(\ref{Popov-propagators}).

The Popov results for point-like bosons (cf. the Appendix) are fully recovered in strong coupling, since in this limit
\begin{eqnarray}
A(q) & \simeq & -  \frac{m^{2} a_{F}}{8 \pi}  
\left[ i \Omega_{\nu}  -  \frac{\mathbf{q}^{2}}{2 m_{B}}  +  \mu_{B}  -  \frac{8 \pi a_{F}}{m}  n_{0} \right]   \nonumber \\
B(q) & \simeq & -  \frac{m^{2} a_{F}}{8 \pi}  \left[ -  \frac{4 \pi a_{F}}{m}  n_{0} \right]                     \nonumber \\
\Sigma_{B}^{\rm{Pop}}(q)_{11} & \simeq & -  \frac{m^{2} a_{F}}{8 \pi}  \left[ \frac{8 \pi a_{F}}{m}  n' \right] \,\, .  \label{Popov-matrix-elements-sc}
\end{eqnarray} 
To get these expressions, Eqs.(\ref{alpha-general-approx}) and 
(\ref{Sigma-Popov-comp-bosons}) have been utilized together with the generic 
definition 
$\mu_{B} = 2 \mu + \epsilon_{0}$ of the bosonic chemical potential in terms of 
the fermionic one.
Use of the specific form (\ref{H-P-Popov}) for $\mu_{B}$, as required by the 
condition (\ref{Gapless-condition-Popov-cb}), then reduces the 
expressions (\ref{Popov-matrix-elements-sc}) to those reported in the 
Appendix (apart from the overall factor $-m^{2} a_{F}/(8 \pi)$).
As a consequence, the expression for $n'$ coincides within the two 
(Bogoliubov and Popov) approximations in strong coupling. 

The final form of the fermionic self-energy is eventually obtained by 
reconsidering the expressions (\ref{fermionic-fluctuations-self-energies}), 
where the Popov propagator $\Gamma^{\rm{Pop}}_{11}(q)$ of Eq.~(\ref{Popov-propagators}) takes the place of $\Gamma_{11}(q)$ and $\Delta$ satisfies the 
condition (\ref{Gapless-condition-Popov-cb}) in the place of the gap equation 
(\ref{general-gap-equation}).
In this way, in the strong-coupling limit the composite bosons are described by the Popov approximation in the place of the Bogoliubov approximation
of Section II.

The last step of our program would require us to modify also the density equation (\ref{fluctuations-density-equation}), consistently with the above 
approximations.
For a homogeneous system, however, no difference would show up in strong coupling from the result obtained in subsection IIC.

All the above results within the Popov approximation for composite bosons are consistent with the relation $a_{B} = 2 a_{F}$.
The same treatment which was made in subsection IIB to improve on this relation (in order to reduce the value of the ratio $a_{B}/a_{F}$ from $2$ to
$0.75$) can be here applied.
What one should do in this case is to first rewrite $\Sigma_{B}^{\rm{Pop}}(q)_{11}$ of Eq.~(\ref{Sigma-Popov-comp-bosons-full}) in terms of the (bare) 
boson-boson interaction (\ref{u-q}) as follows
\begin{equation}
\Sigma_{B}^{\rm{Pop}}(q)_{11} \, \simeq \, - \, 2 \, \int \! \frac{d \mathbf{q'}}{(2\pi)^{3}} \, \frac{1}{\beta} \, \sum_{\nu'} \,
\bar{u}_{2}(q',q,q',q) \, \Gamma_{11}(q') \,\, ,                                                           \label{Sigma-Popov-comp-bosons-full-u-2}
\end{equation}
and then replace $\bar{u}_{2}$ herein by the t-matrix $\bar{t}_{B}$ for the composite bosons of Eq.~(\ref{bosonic-t-matrix}) keeping the four-vector
arguments of Eq.~(\ref{Sigma-Popov-comp-bosons-full-u-2}).
The resulting bosonic self-energy then adds to the diagonal components of Eq.~(\ref{Bogoliubov-self-energy-cb}) when the corresponding replacement is
made.
The Popov propagator for the composite bosons is then obtained as in subsection IIB via the Dyson's equation, in a similar fashion to the procedure for 
point-like bosons described in the Appendix.
Correspondingly, the gapless condition is given by Eq.~(\ref{HP-Bogoliubov-cb}) where now the diagonal self-energy includes also the Popov contribution
with $\bar{t}_{B}$.
This condition reduces again to the gap equation (\ref{Delta-general-equation}) in the strong- and weak-coupling limits (albeit with the modified value
of $a_{B}$ in the strong-coupling limit), in analogy to what was already shown for Eq.~(\ref{Gapless-condition-Popov-cb}).

It is worth emphasizing that, similarly to the analogous treatment of subsection IIB within the Bogoliubov approximation, the price to pay for including
the boson-boson scattering processes beyond the Born approximation is to give up the self-consistency of the fermionic 
single-particle Green's functions entering the ladder propagators (\ref{Gamma-propagators}).
This self-consistency is actually irrelevant in the strong-coupling limit, since it introduces subleading fermionic processes which do not map 
in that limit into corresponding processes for the composite bosons, in analogy to what was shown to occur in the normal phase \cite{PS-2000}.

\section{Concluding remarks}

In this paper, a diagrammatic theory for the BCS-BEC crossover has been set up for the constituent fermions, that treats the composite bosons which 
form in strong coupling in analogy to the Popov approximation for point-like bosons, with mass $m_{B}=2m$, chemical potential 
$\mu_{B} = 2 \mu + \epsilon_{0}$, and coupling constant $4 \pi a_{B}/m_{B}$.
Care has also been taken to include diagrammatic terms which modify 
the value of the boson-boson scattering length $a_{B}$ from the Born result
$a_{B}=2a_{F}$, following the procedure introduced in 
Ref.~\onlinecite{PS-2000}.

A method of successive improvements of the fermionic diagrammatic theory has been presented, according to which the Dyson's equations for the 
constituent fermions and for the composite bosons have been treated at progressive levels of sophistication, in an open-ended fashion.
At any step, physical quantities (like the density) which can be calculated directly from the fermionic single-particle Green's functions 
then include more physical processes than the propagator for the composite bosons entering the fermionic self-energy.
It is thus clear that the relation between conserving and gapless approximations for composite bosons, which was formally established in Ref.~\onlinecite{SP-2004}, 
does not apply by construction to the present scheme of successive approximations.
Nevertheless, it is significant that the bosonic self-energy of Fig.~4 results from taking partially into account the dressing of the fermionic 
single-particle lines within the self-consistent fermionic t-matrix approximation.
It can also be shown that additional diagrams for the effective interaction in the particle-particle channel, which are required to implement the conserving
nature of this fermionic t-matrix approximation in this channel \cite{SP-2004}, can be associated with (at least some of) the boson-boson scattering processes 
beyond the Born approximation discussed in Ref.~\onlinecite{PS-2000}.

In the proposal presented in this paper, we have required the composite-boson propagator entering the fermionic self-energy in the broken-symmetry phase
to be gapless.
Such an approximation, however, cannot be conserving at the same time.
The reason for this choice is that for the BCS-BEC crossover it is important to control the many-body approach in the two BCS and BEC limits, where the small
parameter $k_{F} |a_{F}|$ is used to control the approximations.
In this spirit, it appears appropriate to guarantee the composite-boson propagator entering the fermionic self-energy to be gapless at any coupling,
while satisfying the conserving criterion only perturbatively in the small parameter.
This approach has to be contrasted with the use of the self-consistent fermionic t-matrix approximation in the broken-symmetry phase to deal with the
BCS-BEC crossover \cite{Haussmann}.
In that approach, while satisfying exactly the conserving criterion at the single-particle level for any coupling, the composite-boson propagator
entering the fermionic self-energy is gapped.
This is because the additional diagrams for the effective interaction in the particle-particle channel, which would be needed to make this propagator gapless
\cite{SP-2004}, are not included when constructing this self-consistent t-matrix self-energy.

It is important to emphasize that the fermionic self-energy of Fig.~4, which was exploited in Section III to build up the Popov approximation 
for the composite bosons from the fermionic two-particle Green's functions, is associated with pseudo-gap phenomena for the constituent fermions, both in 
the normal \cite{PPSC-2002} and broken-symmetry \cite{PPS-2004,PPS-PRB-04} phases.
In this way, the refinements of the Bogoliubov approximation for the composite bosons, which led to the Popov approximation, require one to include 
pseudo-gap phenomena not only in the (fermionic) single-particle Green's function but also in the two-particle Green's function.
In a related fashion, as pseudo-gap phenomena evolve with continuity across the critical temperature, the Popov theory for the composite bosons  
in the broken-symmetry phase similarly evolves into the theory of dilute interacting composite bosons in the normal phase \cite{PS-2000}.

The Popov approximation in the broken-symmetry phase is known 
to be valid~\cite{Popov} for 
temperatures at which the condensate ($n_{0}$) and noncondensate
($n'$) densities are comparable to each other and to the total density 
$n$.  
This excludes the temperature ranges close to zero temperature and to the critical temperature, where the Popov approximation may produce
incorrect results.
Further improvements beyond the Popov approximation are accordingly required
in these temperature ranges.

The theoretical framework presented in this paper for the BCS-BEC crossover within the Popov approximation for the composite bosons awaits 
implementation by explicit numerical calculations of physical quantities, similarly to what was recently done within the Bogoliubov approximation for the 
composite bosons, both for a homogeneous system \cite{PPS-2004,PPS-PRB-04} and for trapped atoms \cite{PPPS-PRL-04}.
At least in strong coupling, differences between the Popov and Bogoliubov approximations should show up only in the presence of a trapping potential,
since for a homogeneous system these approximations differ only as far as the value of the bosonic chemical potential is concerned.
Work along these lines is in progress.
Numerical work should, in particular, assess the deviations from the Popov approximation for point-like bosons due to the composite nature of the bosons when approaching 
the intermediate-coupling region, an issue which can be subject to experimental verification.
In this context, one may preliminary test to what an extent the occurrence of a downward shift in the critical temperature of point-like bosons treated
within the Popov approximation \cite{GPS-1996} persists when the composite nature of the bosons begins to matter.

\acknowledgments

This work was partially supported by the Italian MIUR under contract Cofin-2003 
``Complex Systems and Many-Body Problems''. 

\appendix
\section{Bogoliubov and Popov approximations for point-like bosons}

For the sake of completeness, in this Appendix we briefly recall the Bogoliubov and Popov approximations for point-like bosons in the homogeneous case.
The results here reported are used in the text to obtain similar descriptions for the composite bosons in the strong-coupling limit of the 
fermionic attraction.

Quite generally, the Dyson's equation for the propagators $\mathbf{G}_{B}(q)$ of point-like bosons in the broken-symmetry phase reads 
(in matrix notation):
\begin{equation}
\mathbf{G}_{B}(q) \, = \, \mathbf{G}_{B}^{(0)}(q) \, + \, \mathbf{G}_{B}^{(0)}(q) \, \mathbf{\Sigma}_{B}(q) \, \mathbf{G}_{B}(q) \,\, . 
\label{boson-Dyson-equation}
\end{equation}
In this expression, $\mathbf{G}_{B}^{(0)}(q)$ is the free-boson propagator with inverse
\begin{equation}
\mathbf{G}_{B}^{(0)}(q)^{-1} = \left( \begin{array}{cc} 
i \Omega_{\nu} -  \frac{\mathbf{q}^{2}}{2 m_{B}} +  \mu_{B} &   0   \\
0 & - i \Omega_{\nu} -  \frac{\mathbf{q}^{2}}{2 m_{B}}  +  \mu_B
\end{array} \right)
\label{inverse-free-boson}
\end{equation}
where $m_{B}$ and $\mu_{B}$ are the bosonic mass and chemical potential, respectively, and $\mathbf{\Sigma}_{B}(q)$ is the $2 \times 2$ bosonic self-energy.

Within the Bogoliubov approximation, $\mathbf{\Sigma}_{B}(q)$ is taken of the form \cite{FW}
\begin{equation}
\mathbf{\Sigma}_{B}^{{\rm Bog}}(q) \, = \, g \, n_{0} \, \left( \begin{array}{cc} 
2 &  1   \\
1 &  2
\end{array} \right)                                                                     \label{Bogoliubov-self-energy}
\end{equation}
where $n_{0}$ is the condensate density.
In this expression, the constant coupling $g = 4 \pi a_{B}/m_{B}$ results from replacing the bare (repulsive) boson-boson potential by a t-matrix, 
which is in turn considered equal to the constant value for the scattering of two bosons in vacuum associated with the scattering length $a_{B}$ 
\cite{footnote-g}.
With the self-energy (\ref{Bogoliubov-self-energy}), the Dyson's equation (\ref{boson-Dyson-equation}) yields:
\begin{eqnarray}
&&\!\!\!\!\!\mathbf{G}_{B}^{{\rm Bog}}(q) = \frac{1}{\Omega_{\nu}^{2} +
 E_{B}(q)^{2}}\nonumber\\
&&\!\!\!\!\times\left( \begin{array}{cc} 
- i \Omega_{\nu} - \frac{\mathbf{q}^{2}}{2 m_{B}} - g n_{0} &  g n_{0}   \\
g n_{0} & i \Omega_{\nu}  -  \frac{\mathbf{q}^{2}}{2 m_{B}}  - g n_{0}
\end{array} \right) 
\label{Bogoliubov-propagators}
\end{eqnarray}
where the relation $\mu_{B} = g n_{0}$ that holds within the Bogoliubov approximation has been used and where
\begin{equation}
E_{B}(\mathbf{q}) = \sqrt{ \left( \frac{\mathbf{q}^{2}}{2 m_{B}}  + g n_{0} 
\right)^{2} - g^{2} \ n_{0}^{2}}
 \label{Bogoliubov-dispersion}
\end{equation}

\noindent
is the (gapless) Bogoliubov quasiparticle dispersion.

The Popov approximation is obtained by considering the additional diagonal self-energy \cite{Popov}
\begin{equation}
\mathbf{\Sigma}_{B}^{\rm{Pop}}(q) = 2 g \, n' \left( \begin{array}{cc} 
1 &  0   \\
0 &  1
\end{array} \right)                                                                     \label{Popov-self-energy}
\end{equation}
where $n'$ is the noncondensate density such that $n_{0} + n'$ is the total bosonic density.
The bosonic propagators $\mathbf{G}_{B}^{\rm{Pop}}(q)$  within the Popov 
approximation can be related to the propagators (\ref{Bogoliubov-propagators}) 
within the Bogolibov approximation via the modified Dyson's equation:
\begin{eqnarray}
\mathbf{G}_{B}^{\rm{Pop}}(q) 
 =  \mathbf{G}_{B}^{\rm{Bog}}(q)  +  
\mathbf{G}_{B}^{\rm{Bog}}(q)  \mathbf{\Sigma}_{B}^{\rm{Pop}}(q) 
\mathbf{G}_{B}^{\rm{Pop}}(q) \,\, .
\label{modified-boson-Dyson-equation}\\
\nonumber
\end{eqnarray}
The relation $\mu_{B} = g (n_{0} + 2 n')$ is required for the propagators to
 be gapless within the Popov approximation.
Exploiting this relation, it can be shown that the Popov propagators 
$\mathbf{G}_{B}^{\rm{Pop}}(q)$ [Eq.~(\ref{modified-boson-Dyson-equation})] 
acquire the \emph{same} form of the Bogoliubov propagators 
$\mathbf{G}_{B}^{\rm{Bog}}(q)$ [Eq.~(\ref{Bogoliubov-propagators})], once the 
respective chemical 
potentials are eliminated in favor of the condensate density $n_{0}$.
For the homogeneous case, the expression for $n'$ thus coincides within the 
two (Bogoliubov and Popov) approximations \cite{footnote-nh}.

The above coincidence between the Bogoliubov and Popov bosonic propagators 
stems from the simplifying assumption that the self-energies
(\ref{Bogoliubov-self-energy}) and (\ref{Popov-self-energy}) contain the 
\emph{constant} coupling $g$, thus omitting the complicated (frequency and
wave-vector) structure of the bosonic many-body t-matrix occurring in the
 formal expressions of the self-energies reported in Ref.\onlinecite{Popov}.


\begin{thebibliography}{99}
\bibitem{footnote-strong-coupling} In the strong-coupling limit of the fermionic attraction, the magnitude $|\mu|$ of the fermionic chemical potential
                                   is the largest energy scale in the problem.
                                   In this limit, the conditions $\Delta \ll |\mu|$ and $\beta \mu \rightarrow - \infty$ can thus be either
                                   separately or simultaneously applied,
 depending on the need.
\bibitem{footnote-nh} In the presence of a trapping potential, it can be shown from general arguments that the functions 
                      $n_{0}(\mathbf{r})$ and $n'(\mathbf{r})$ obtained within the Popov approximation do not coincide with 
                      those obtained within the Bogoliubov approximation.
\end{thebibliography}
\end{document}